\documentclass[10pt]{iopart}

\usepackage{lineno,hyperref}
\modulolinenumbers[5]

\usepackage[utf8]{inputenc}
\usepackage[english]{babel}
\usepackage[export]{adjustbox}
\usepackage{xcolor}
\usepackage{multirow}

\usepackage{graphicx}
\usepackage{ulem} 

\usepackage{hyperref}
\usepackage[all]{hypcap} 

\usepackage{bm}
\usepackage{tabularx}
\usepackage{longtable}

\begin{document}

\title{Enhanced magnetism and suppressed magnetoelastic coupling induced by electron doping in Ca$_{1-x}$Y$_{x}$MnReO$_6$}

\author{A S Cavichini$^1$\footnote{Corresponding author: \ead{cavichiniart@gmail.com}}, M T Orlando$^1$, M C A Fantini$^2$, R Tartaglia$^3$, C W Galdino$^3$, F Damay$^4$, F Porcher$^4$, E. Granado$^3$}

\address{$^1$Universidade Federal do Esp\'irito Santo, Vit\'oria, Esp\'irito Santo 29075-910, Brazil}
\address{$^2$Institute of Physics, University of S\~ao Paulo, S\~ao Paulo, 05508-090,  Brazil}
\address{$^3$"Gleb Wataghin'' Institute of Physics, University of Campinas - UNICAMP, Campinas, S\~ao Paulo 13083-859, Brazil}
\address{$^4$Laboratoire L\'eon Brillouin, CEA, Centre National de la Recherche Scientifique, CE-Saclay, 91191 Gif-sur-Yvette, France}

\begin{abstract}

The Ca$_2$MnReO$_6$ double perovskite is a spin-orbit-assisted Mott insulator with exotic magnetic properties, including a largely non-collinear Mn$^{2+}$ spin arrangement and nearly orthogonal coupling between such spins and the much smaller Re $5d$ magnetic moments. Here, the electron-doped compound Ca$_{1.7}$Y$_{0.3}$MnReO$_6$ is investigated. Neutron and X-ray powder diffraction confirm that nearly full chemical order is maintained at the Mn and Re sites under the Y substitution at the Ca site. X-ray absorption measurements and an analysis of the Mn-O/Re-O bond distances show that the Mn oxidation state remains stable at +2 whereas Re is reduced upon doping. The electron doping increases the magnetic ordering temperature from $T_c = 121$ to $150$ K and also enhances significantly the ferromagnetic component of the Mn spins at the expense of the antiferromagnetic component at the base temperature ($T=3$ K). The lattice parameter anomalies at $T_c$ observed in the parent compound are suppressed by the electron doping. The possible reasons for the enhanced magnetism and the suppressed magnetoelastic coupling in Ca$_{1.7}$Y$_{0.3}$MnReO$_6$ are discussed.

\end{abstract}


\submitto{\JPCM}

\maketitle


\section{Introduction}

In materials containing heavy transition-metal ions in the $4d$ and $5d$ rows, the presence of spin-orbit interactions, crystal-field splittings and electronic correlations on equivalent energy scales may lead to exotic spin-orbital ground states with surprising interrelated electronic, magnetic and structural behavior \cite{Cao2018}. The recent spark of interest on $4d$ and $5d$-based magnetism naturally leads to pursuing materials presenting both $3d$ and $5d$ ions, where non-trivial effects arising from the coexistence and interactions of $3d$ and $5d$ localized moments may take place \cite{Bufaical2016}. Perhaps the simplest structural motif that allows for such investigations is the double perovskite structure with alternating $M_{3d}$O$_6$ and $M_{5d}$O$_6$ octahedra.

Recently, the magnetism of the Ca$_2$MnReO$_6$ (CMRO) monoclinic double perovskite was investigated, showing dominant noncollinear Mn magnetic moments [4.35(7) $\mu_B$] with a nearly orthogonal alignment with respect to the small Re moments [0.22(4) $\mu_B$] (Ref. \cite{Cavichini2018}), such as previously found in the similar material Sr$_2$MnReO$_6$ (Ref. \cite{Popov2003}). CMRO was also recognized, on the basis of Density Functional Theory (DFT) calculations, as a spin-orbit-assisted Mott insulator \cite{Cavichini2018}, where the states nearby the Fermi level have a mixed Re $5d$ and O $2p$ character, with no contribution from the Mn $3d$ states. Our initial study on pure CMRO led to questions about the possible effects of chemical doping on the magnetism and electronic properties of this system. In this work, we investigate the crystal and magnetic structures of Ca$_{1.7}$Y$_{0.3}$MnReO$_6$ (CYMRO) by means of neutron powder diffraction (NPD), X-ray powder diffraction (XPD), X-ray absorption spectroscopy (XAS), and dc-resistivity. We find that the electron doping associated with the partial Ca$^{2+}$ substitution for Y$^{3+}$ reduces the average Re oxidation state without altering the pure Mn$^{2+}$ valence as expected by previous DFT calculations \cite{Cavichini2018}, however without inducing metallization. The magnetic structure of CYMRO remains largely non-collinear with a much larger ferromagnetic (FM) component of the dominating Mn moments along the monoclinic principal axis $b$-direction with respect to the parent compound CMRO. The magnetic ordering transition temperature $T_c$ increases substantially with Y doping. Our observations are rationalized in terms of the presence of Re $5d$ $J_{eff}=3/2$ states with substantial spin and orbital magnetic moments that nearly cancel each other in CMRO. The extra Re electronic density brought by the Y substitution appears to enhance the Mn-O-Re-O-Mn superexchange interaction and disrupt the Re orbital coherence. This system offers a rare opportunity to investigate the magnetic interactions between localized $3d$ and $5d$ moments.

\begin{figure*}[h]
\centering
\includegraphics[width=0.8\textwidth]{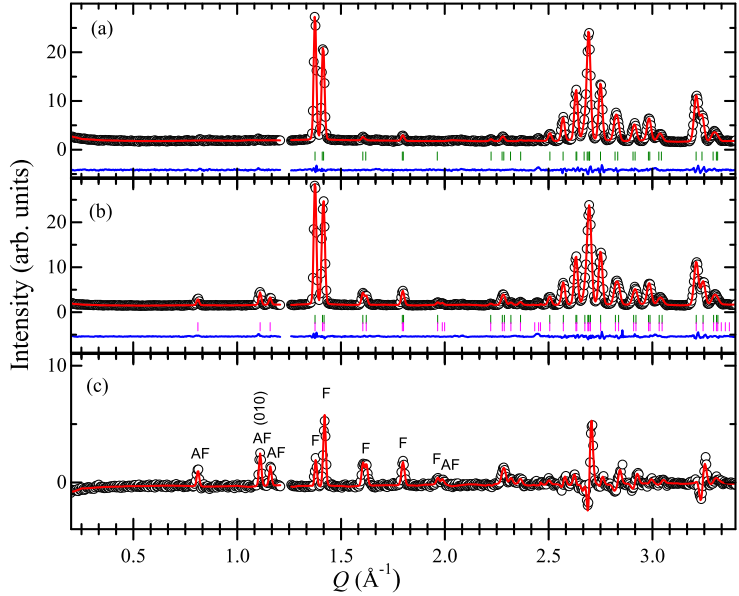}
\caption{\label{c-npd} Observed cold neutron powder diffraction profiles of Ca$_{1.7}$Y$_{0.3}$MnReO6 at $T=155$ K (a) and $T=3$ K (b) (open circles). The difference profile $I$(3 K)-$I$(155 K) is shown in (c). The calculated profiles for the refined nuclear ($T=155$ K) and nuclear+magnetic ($T=3$ K) structures are shown as red solid lines. The difference curves between observed and calculated profiles are displayed at the bottom of (a) and (b) (blue solid lines). The positions of the symmetry-allowed Bragg reflections are indicated as short green and magenta vertical bars for the nuclear (a and b) and magnetic (b) structures, respectively. In (c), the magnetic reflections arising from ferromagnetic (F) and antiferromagnetic (AF) components are explicitly indicated for $Q<2$ \AA$^{-1}$.}
\end{figure*}

\section{Experimental Details}

Polycrystalline Ca$_{2-x}$Y$_x$MnReO$_6$ ($x=0.0$ , 0.1, 0.2 and 0.3) samples were synthesized by solid state reaction through the same synthesis route as in our previous work with CMRO \cite{Cavichini2018,Depianti2013} with addition of stoichiometric Y$_2$O$_3$ (Sigma Aldrich, 99.995\%) quantities to achieve the expected composition. The phase purity was verified for each sample by X-ray powder diffraction. {\it ac}-Magnetic susceptibility measurements indicate a monotonic increment of $T_c$ from $\sim 120$ K for $x=0.0$ to $\sim 150$ K for $x=0.3$ [see Appendix], thus the $x=0.3$ compound was selected for a more detailed investigation. {\it dc}-Electrical resistivity measurements indicate a non-metallic behavior for $x=0.3$ that is well described by the Efros and Shklovski variable-range hopping \cite{Efros_1975} approach (see SM) .

Energy-dispersive X-ray Absorption Spectroscopy (XAS) measurements were performed at the Mn $K$ and Re $L_3$ edges at the DXAS beamline of the Brazilian synchrotron laboratory (LNLS), using a Si(111) curved crystal as the polychromator.

Cold neutron powder diffraction (c-NPD) measurements were taken at the $G$4-1 instrument of Laboratorie L\'eon Brillouin (LLB), using a highly oriented pyrolytic graphite (HOPG) monochromator with vertical focusing and $\lambda = 2.43$ \AA, a two-axis diffractometer and a BF$_3$ multicell detector with $80^{\circ}$ aperture and $0.1^{\circ}$ resolution, using $\sim 10$ g of pelletized sample. High-resolution thermal neutron powder diffraction (t-NPD) measurements were taken at selected temperatures at the $3T2$ instrument of LLB using a vertically focusing Ge(335) monochromator with $\lambda = 1.2292$ \AA\ and a bank of 50 $^{3}$He detectors. In both experiments, the sample was kept sealed under He atmosphere in a vanadium can and mounted into the cold finger of a LHe cryostat. The powder diffraction data were analysed using the \textit{FullProf} \cite{RodriguezCarvajal1993} suite. The Re magnetic form factor was taken from Ref. \cite{Popov2003}.  

The Mn/Re antisite disorder of CYMRO was consistently found to be below 1 \% from Rietveld refinements using either c-NPD, t-NPD, or XPD data. The possibility of Y$^{3+}$ ions being incorporated at the Re site was also investigated using XPD data, where the Y ions were allowed to occupy both the Ca and Re sites in the structural refinement. The refined $A$-site Ca/Y occupations are 1.686(10)/0.314(10), very close to the stoichiometric value, whereas the the Y occupation at the Re site is 0.016(8), i.e., near zero within 2 standard deviations. Thus, CYMRO may be described as a $B$-site ordered Mn/Re double perovskite, with the Y substitution being restricted to the Ca site.

\section{Results and Analysis}

\subsection{Neutron diffraction}

\subsubsection{Nuclear structure}

The c-NPD profiles of CYMRO at $T=155$ K ($>T_c$) and $T=3$ K ($<<T_c$) are given in Figures. \ref{c-npd}(a) and \ref{c-npd}(b), respectively. The t-NPD profile at selected temperatures and the XPD profile at room temperature are shown in the SM. The crystal structure of CYMRO could be well modeled by a monoclinic double perovskite structure with space group $P2_1/n$ at all temperatures [see red solid line in Figure \ref{c-npd}(a)], in line with previous work for the parent material \cite{Cavichini2018}. 

Table \ref{struct} shows the refined structural parameters and Table \ref{bonds} displays the Mn-O and Re-O bond lengths and Mn-O-Re bond angles obtained from the data of Table \ref{struct}.  

The average Mn-O distance for CYMRO at $T=300$ K is 2.137(2) \AA, which is very close to the value obtained for CMRO [2.142(2) \AA]. Bond valence sum calculations using the data of Table \ref{bonds} and Ref. \cite{Cavichini2018} indicate a pure Mn$^{2+}$ oxidation state for both CMRO and CYMRO. On the other hand, the average Re-O distance for CYMRO at $T=300$ K is 1.940(2)\AA, substantially larger than for CMRO [1.912(2) \AA] (Ref. \cite{Cavichini2018}). This is consistent with a reduction of the Re oxidation state for CYMRO with respect to the formal Re$^{6+}$ state of pure CMRO, indicating that the Y-substitution in this system effectively dopes the Re-O network with electrons. Further evidence supporting this conclusion is presented in Section III.E below.

\begin{figure}[ht]
\centering
\includegraphics[width=0.5\textwidth]{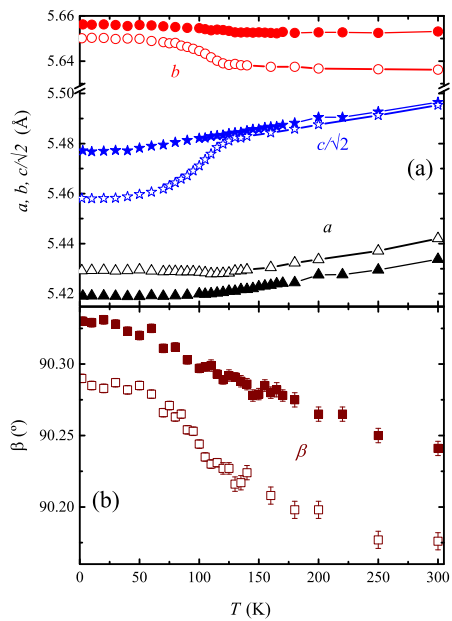}
\caption{\label{Tdep_lattice} Temperature dependence of the (a) lattice parameters and (b) $\beta$ angle of Ca$_{1.7}$Y$_{0.3}$MnReO$_6$ (filled symbols) in comparison to the parent compound Ca$_2$MnReO$_6$ (open symbols, taken from Reference \cite{Cavichini2018}). The statistical errors in (a) are not displayed for being smaller than the symbol size.} 
\end{figure}

Figures \ref{Tdep_lattice}(a) and \ref{Tdep_lattice}(b) show the temperature-dependence of the refined lattice parameters, in comparison to previously published data for CMRO  \cite{Cavichini2018}. We note that the lattice anomalies below $T_c=121$ K for CMRO, associated with the magnetic ordering temperature, are nearly vanished for CYMRO. An analysis of the bond distances indicates that the lattice anomalies for CMRO below $T_c$ are associated with a contraction of Re-O$_3$ with respect to Re-O$_1$ and Re-O$_2$ bond lengths \cite{Cavichini2018}, leading to a more distorted ReO$_6$ octahedra below $T_c$ for the pure compound, which is not noticed for CYMRO (see Table \ref{bonds}).

\begin{table*}
\centering
\caption{\label{struct} Structural parameters, atomic coordinates and $R$ factors of CYMRO (sample $2$) at three distinct temperatures using thermal neutron powder diffraction data. B-site fixed coordinates are Mn/Re=(0.5/0.5 0/0 0.5/0). }
\begin{indented}
\item[]\begin{tabular}{c c c c c} 
\br
& 10 K & 150 K & 300 K \\
\mr
$a$ ({\AA}) & 5.41855(9) & 5.43072(11) & 5.43272(8) \\
$b$ ({\AA}) &5.65080(9) &5.63810(11) & 5.65480(8) \\
$c$ ({\AA}) & 7.74673(12) & 7.75741(16) & 7.7731(1) \\
$\beta$ ({\AA}) & 90.3217(12) & 90.2204(14) & 90.2349(11) \\
Cell Vol. {(\AA$^3$}) & 237.458(7)  & 237.522(8) & 238.795(6) \\
\mr
Ca/Y  \\
x & 0.4833(5) & 0.4879(9) & 0.4849(6) \\
y & 0.5601(6) & 0.5541(7) & 0.5579(5) \\
z & 0.2556(5) &0.2556(6) & 0.2535(4) \\
B ({\AA$^2$}) & 0.61(4) & 0.83(7) & 1.12(5) \\
Mn/Re  \\
B ({\AA$^2$}) &0.67(7)/0.23(3) & 0.27(9)/0.22(5) & 0.83(7)/0.49(3) \\
O1  \\
x & 0.3170(4) & 0.3179(7) &0.3163(5) \\
y & 0.2833(4) & 0.2818(6) & 0.2839(4) \\
z & 0.0575(3) & 0.0545(5) &0.0566(4) \\
B ({\AA$^2$}) & 0.65(5) & 0.66(5) & 1.10(5) \\
O2 \\
x & 0.2069(6) &0.2092(6) & 0.2067(5) \\
y & 0.8120(4) & 0.8133(6) & 0.8137(4) \\
z & 0.0480(4) & 0.0465(5) &0.0476(4) \\
B ({\AA$^2$}) & 0.75(5) & 0.30(5) & 1.07(5) \\
O3  \\
x & 0.6031(4) &0.5968(7) & 0.6012(4) \\
y & -0.0364(5) & -0.0323(7) & -0.0352(5) \\
z & 0.2370(3) & 0.2351(5) & 0.2371(3) \\
B ({\AA$^2$}) & 0.76(4) & 0.76(6) & 1.00(4) \\
\mr
$R_p$ (\%) & 2.43 & 3.29 & 2.23 \\
$R_{wp}$ (\%) & 3.16 & 4.39 & 2.90 \\
$\chi^2$ & 5.55 & 6.56 & 6.75 \\
\br
\end{tabular}
\end{indented}
\end{table*}

\begin{table}
\centering
\caption{\label{bonds} M-O bond distances and Mn-O-Re bond angles for CYMRO (sample $2$) extracted from data of Table \ref{struct}.}
\begin{indented}
\item[]\begin{tabular}{c c c c c} 
\br
& 10 K & 150 K & 300 K \\
\mr

Mn-O$_1$ (${\AA}$) & 2.155(2) & 2.160(4) & 2.153(4) \\
Mn-O$_2$ (${\AA}$) & 2.123(2) & 2.130(3) & 2.131(2) \\
Mn-O$_3$ (${\AA}$) &2.126(2) &2.131(4) & 2.128(2) \\
Re-O$_1$ (${\AA}$) & 1.938(2) & 1.919(4) & 1.942(2) \\
Re-O$_2$ (${\AA}$) & 1.949(3) & 1.933(3) & 1.947(3) \\
Re-O$_3$ (${\AA}$) & 1.927(2) & 1.905(4) & 1.931(2) \\
\mr
Mn-O$_1$-Re ($^{\circ}$) & 146.24(9) & 147.17(14) & 146.48(10) \\
Mn-O$_2$-Re ($^{\circ}$) & 148.26(1) & 148.82(13) & 148.05(10) \\
Mn-O$_3$-Re ($^{\circ}$) & 145.74(9) & 147.88(16) & 146.43(9) \\
\br
\end{tabular}
\end{indented}
\end{table}

\begin{figure}[ht]
\centering
\includegraphics[width=1\textwidth]{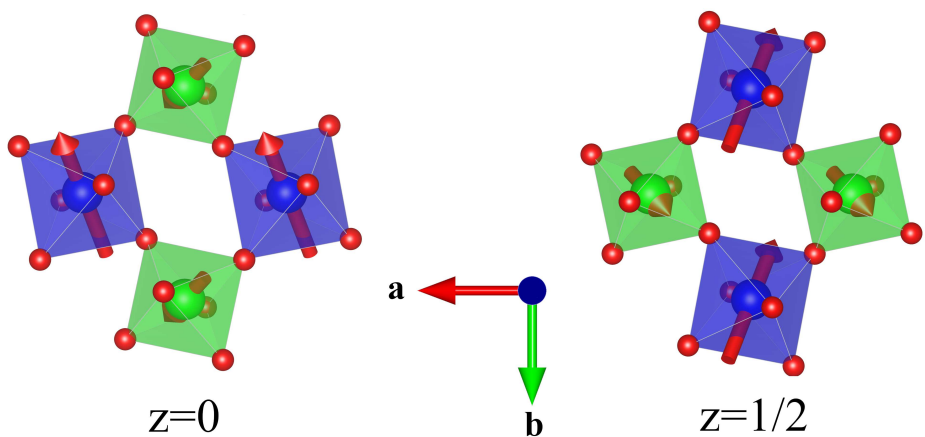}
\caption{\label{crystal} Magnetic structure of Ca$_{1.7}$Y$_{0.3}$MnReO$_6$ showing separately the Mn and Re moments at $z=0$ and $z=1/2$. The magnitude of the weak Re moment is exaggerated for clarity.}
\end{figure}

\begin{figure}[ht]
\centering
\includegraphics[width=0.5\textwidth]{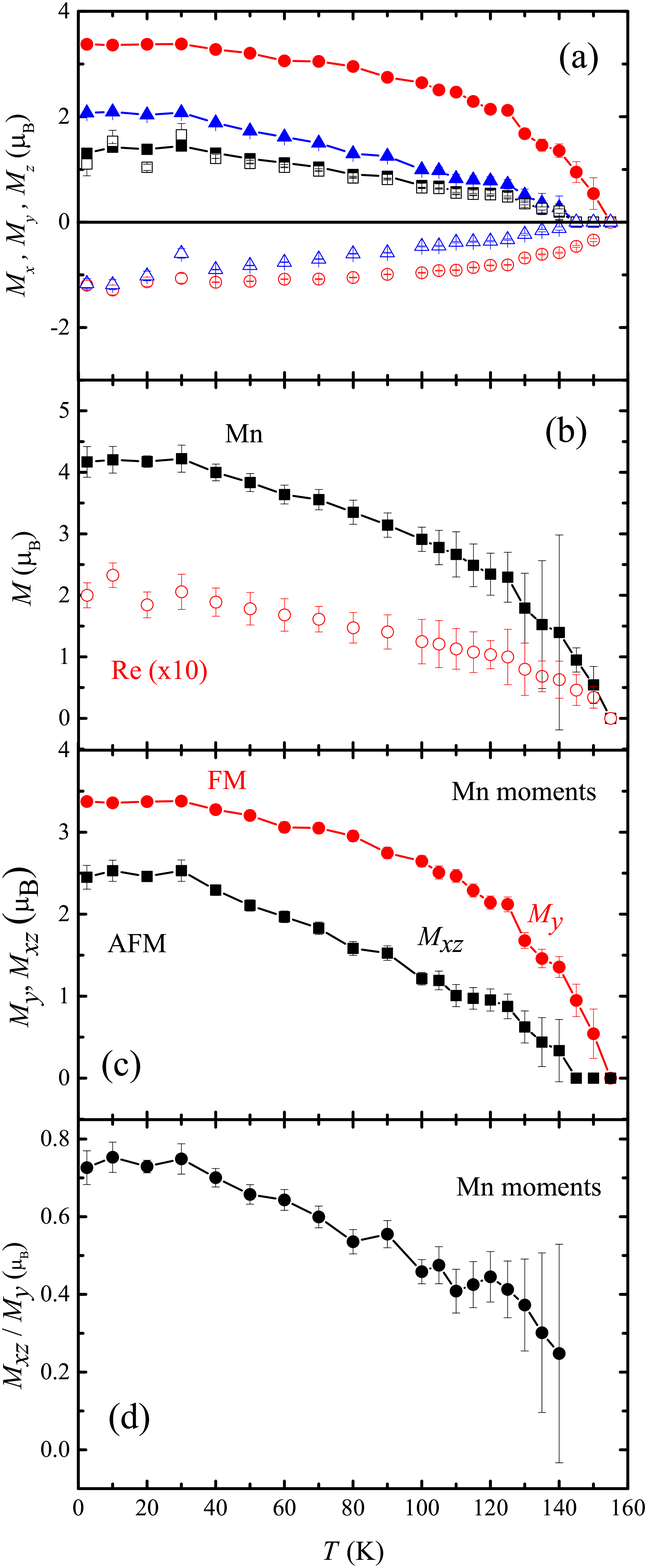}
\caption{\label{Tdep} Temperature dependence of (a) cartesian components of the atomic moments $M_x$ (Mn) (filled squares), $M_y$ (Mn) (filled circles), $M_z$ (Mn) (filled triangles), $M_x$ (Re)  $\times 10$ (open squares), $M_y$ (Re) $\times 10$ (open circles) and $M_z$ (Re) $\times 10$ (open triangles), (b) total Mn and Re ($\times10$) moments, (c) FM (b axis) and AFM (ac plane) components of the Mn moment, and (d) AFM/FM ($M_{xz}/M_y$) ratio of the Mn moments for CYMRO (sample $2$).}
\end{figure}

\subsubsection{Magnetic structure}

Figure \ref{c-npd}(c) shows the intensity difference $I$(3 K) - $I$(155 K) of the c-NPD data for CYMRO. Most of this difference arises from magnetic diffraction at relatively low neutron momentum-transfers ($Q < 2$ \AA $^{-1}$). In opposition, at higher $Q$ the Mn and Re magnetic form factors are substantially reduced, and Bragg peak shifts and nuclear intensity changes due to lattice expansion and atomic displacements with $T$ (see Table \ref{struct}) become increasingly more influential for the neutron intensity difference in Figure \ref{c-npd}(c). It is seen at Figures. \ref{c-npd}(b) and \ref{c-npd}(c) that additional Bragg peaks that are forbidden for the nuclear crystal structure, as well as additional intensities for pre-existing peaks, are observed at low temperatures and low $Q$ due to the long-range magnetic order. In similarity to CMRO (Ref. \cite{Cavichini2018}), all extra contributions of CYMRO can be indexed with a monoclinic magnetic unit cell that coincides with the nuclear cell, which means a $\vec{k}$ = (0,0,0) magnetic structure. Thus, the symmetry-allowed magnetic structures considering Mn/Re occupying 2\textit{c}/2\textit{d} sites (see Table \ref{bonds} for atomic positions in the ITA-setting) were determined by combining representational analysis, using Basireps, a computing program of the \textit{FullProf} suite \cite{RodriguezCarvajal1993}, and Magnetic Space groups (MSG) formalism, using the Bilbao Crystallographic Server \cite{Perez-Mato2015}. The magnetic representation $\Gamma_{mag}$ can be decomposed in the following manner: $$\Gamma^{(Mn)}_{mag} = 3 \Gamma^{1}_{1} \oplus 0 \Gamma^{1}_{2} \oplus 3 \Gamma^{1}_{3} \oplus 0 \Gamma^{1}_{4} $$ and $$\Gamma^{(Re)}_{mag} = 3 \Gamma^{1}_{1} \oplus 0 \Gamma^{1}_{2} \oplus 3 \Gamma^{1}_{3} \oplus 0 \Gamma^{1}_{4} $$ with all the space group irreducible representations (\textit{irreps}) being one-dimensional as indicated by the superscript. From the decomposition, only $\Gamma^{1}_{1}$ and $\Gamma^{1}_{3}$ can generate magnetic structures for both sites. Since there is only one magnetic phase transition, the magnetic atoms should order within the same magnetic representation, which must appear simultaneously in the decomposition of the two magnetic atoms. (either $\Gamma^{1}_{1}$ or $\Gamma^{1}_{3}$ in this case). In Table \ref{irrep} we present the basis vectors for each \textit{irrep} and its Magnetic Space Group.

	\begin{table}[]
\centering
		\caption{Irreducible representations $\Gamma$ and its notation in the Bilbao Crystallographic Server \cite{Perez-Mato2015} with the basis vectors at Mn/Re sites.} 
		\begin{indented}
		\item[]\begin{tabular}{l l l l} 
			\br
			$\Gamma$& Mn$_{1}$/Mn$_{2}$ & Re$_{1}/$Re$_{2}$ & MSG\\
			\mr
	    	\multirow{2}{*}{$\Gamma_{1}$ (mGM$_{1}^{+}$)} & ($m_{x}$,$m_{y}$,$m_{z}$) & ($m_{x}$,$m_{y}$,$m_{z}$), & \multirow{2}{*}{$P2_{1}/n$} \\  
		 & (-$m_{x}$,$m_{y}$,$-m_{z}$) & (-$m_{x}$,$m_{y}$,$-m_{z}$) \cr \cr \\ 
		\multirow{2}{*}{$\Gamma_{3}$ (mGM$_{2}^{-}$)} & ($m_{x}$,$m_{y}$,$m_{z}$), & ($m_{x}$,$m_{y}$,$m_{z}$) &  \multirow{2}{*}{$P2_{1}$$'$$ /n'$} \\   & ($m_{x}$,-$m_{y}$,$m_{z}$) & ($m_{x}$,-$m_{y}$,$m_{z}$)  \cr  \cr 
		
			\br 
		\end{tabular}
		\label{irrep} 
	\end{indented}	
	\end{table}

The paramagnetic $P2_1/n$ space group allows for two magnetic space groups ($P2_1/n$ and $P2_{1}$$'$$ /n'$), both accommodating non-collinear order that may be decomposed into FM and AFM components for each magnetic ion (see Table \ref{irrep}). For the $P2_1/n$ magnetic group, the AFM component lies in the {\it ac} plane and the FM component is directed along the \textit{b} direction. For $P2_{1}$$'$$ /n'$, the inverse happens, i.e., the AFM component is directed along \textit{b} and the FM component lies in the {\it ac} plane. The observation of the (010) AFM reflection [see Figure \ref{c-npd}(c)] rules out the $P2_{1}$$'$$ /n'$ group for CYMRO, thus the refinements at low temperatures shown as solid lines in Figures. \ref{c-npd}(b) and \ref{c-npd}(c) were performed using the basis vectors of $P2_1/n$ magnetic group. Despite the substantial number of magnetic fitting parameters (three independent components for each magnetic ion, six in total), the refinements at the base temperature successfully converged without the necessity of constraining the fitting parameters, leading to physically acceptable refined moments. The thus-obtained magnetic structure is displayed in Figure \ref{crystal}. At $T=3$ K, the refined Mn/Re moment vectors are $[1.31(7), 3.38(3), 2.07(3)]$/$[-0.11(7) ,-0.12(3),0.12(3)]$ $\mu_B$, yielding total Mn/Re magnetic moments of $4.18(7)/0.20(4)$ $\mu_B$. The canting angle between Mn and Re moments is 110(5)$^{\circ}$ for ions located at the same $z$ and 130(20)$^{\circ}$ for ions in consecutive planes. Finally, the Mn-Mn and Re-Re magnetic canting angles for atoms in consecutive planes are 59.5(5)$^{\circ}$ and 110(20)$^{\circ}$, respectively.

On warming towards the magnetic ordering temperature of CYMRO, the magnetic intensities of the Bragg peaks obviously become smaller, and as a consequence the convergence of the fit with the six unconstrained magnetic parameters become more challenging. Thus, in order to analyze the temperature-dependence of the Mn and Re magnetic moments, we fixed the $M_x$/$M_z$ ratio for both Mn and Re moments at the values obtained at the base temperature (see above), which prevented the refinements at finite temperatures from diverging.
The so-obtained temperature dependence of the refined Mn and Re moments components are shown in Figure \ref{Tdep}(a), whereas the total moments are given in Figure \ref{Tdep}(b). Figure \ref{Tdep}(c) shows the temperature-dependencies of FM ($M_y$) and AFM ($M_{xz}$) components of the Mn moment, revealing distinct order parameters for these components. The $M_{xz}/M_y$ ratio is given in Figure \ref{Tdep}(d), starting at $\sim 0.75$ at the base temperature and reducing to $\sim 0.3$ at $T=140$ K.

\subsection{X-ray absorption spectroscopy (XAS)}

\begin{figure}[ht]
\centering
\includegraphics[width=0.8\textwidth]{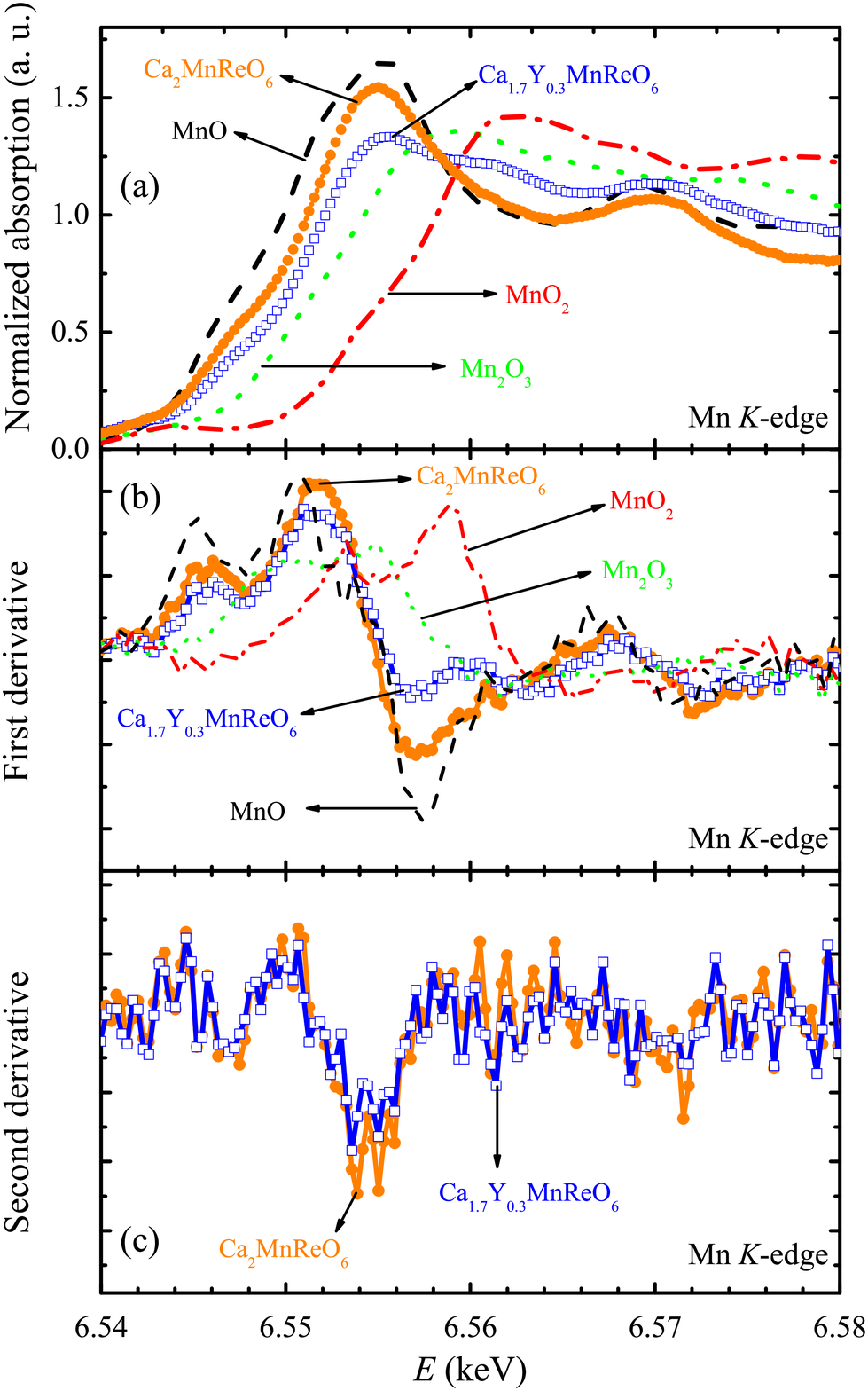}
\caption{\label{XANES-Mn} (a) X-ray Absorption Near Edge Structure (XANES) at the Mn $K$ edge of CMRO (closed circles), CYMRO (open circles), and the reference materials MnO (dashed line), Mn$_2$O$_3$ (dotted line), MnO$_2$ (dash-dotted line) and metallic Mn (solid line). (b) First and (c) second derivatives XANES spectrum of CMRO and CYMRO at the Mn $K$ edge.}
\end{figure}

\begin{figure}[ht]
\centering
\includegraphics[width=0.8\textwidth]{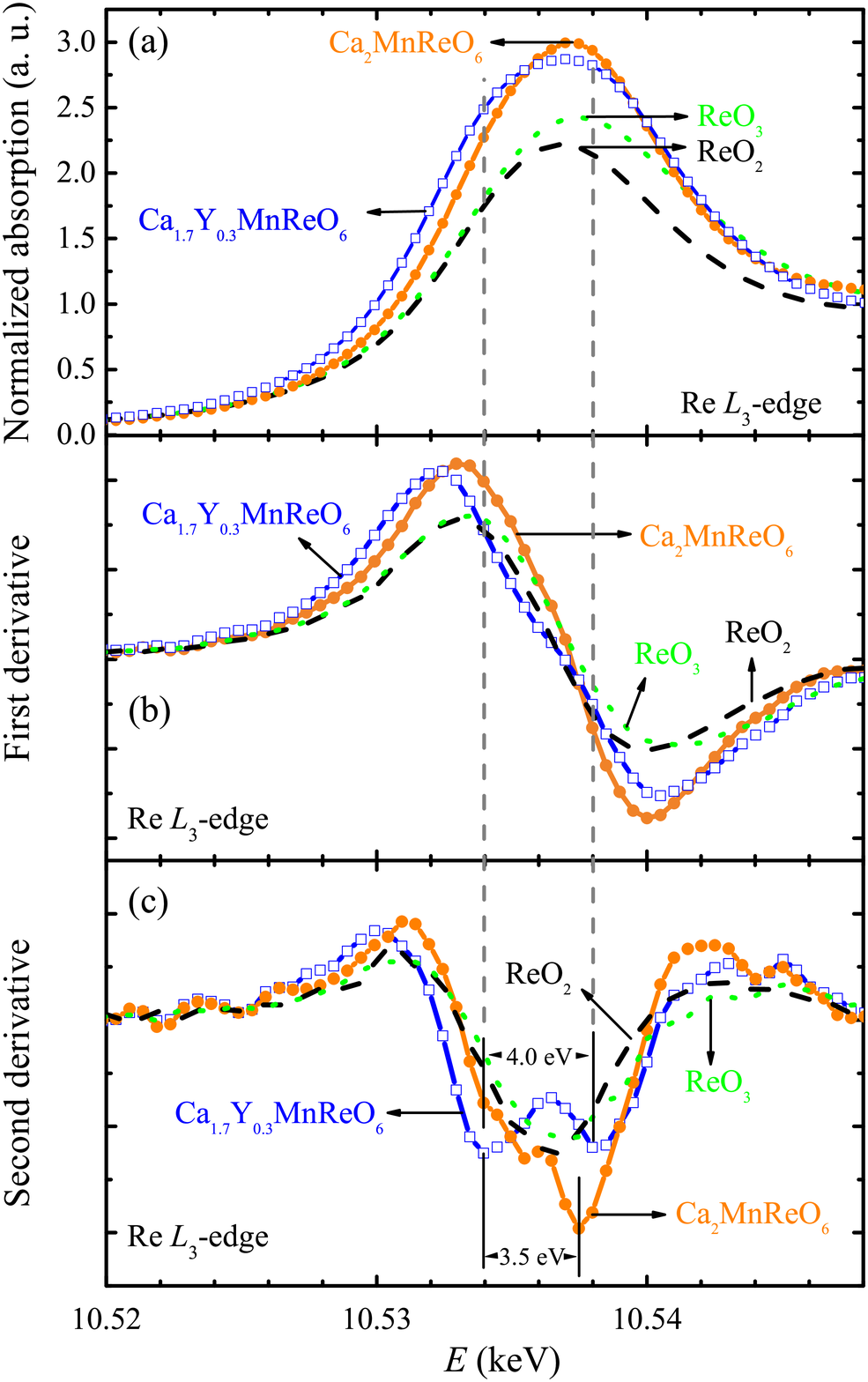}
\caption{\label{XANES-Re} (a) Re $L_3$ XANES of CMRO, CYMRO (symbols), and the reference materials ReO$_2$ (dashed line) and ReO$_3$ (dotted line). (b) First and (c) second derivatives XANES spectrum of CMRO and CYMRO at the Re $L_3$ edge.}
\end{figure}

X-ray Absorption Near Edge Structure (XANES) measurements of CMRO and CYMRO at the Mn $K$ edge are shown in Figure \ref{XANES-Mn}(a) together with MnO, Mn$_2$O$_3$ and MnO$_2$ reference standards. The spectrum of CMRO shows a shoulder at 6.547 keV, a prominent white line at 6.555 keV and another maximum at $\sim 6.57$ keV, consistent with an oxidation state close to Mn$^{2+}$. Although these same spectral features are still observed for CYMRO, significant changes are noticed for this sample with respect to CMRO, most notably a transfer of spectral weight from the white line at 6.555 keV to higher energies, including the formation of a shoulder at $\sim 6.56$ keV that is not clearly observed for CMRO. Still, the Mn \textit{K} edge position, defined as the first maximum of the derivative spectrum  [see Figure \ref{XANES-Mn}(b)] does not show significant variations between CMRO and CYMRO. Also, the second derivative curves for CMRO and CYMRO are identical within our resolution [see Figure \ref{XANES-Mn}(c)].

Analysis of the Mn-O distances obtained from our neutron powder diffraction data indicate a stable Mn$^{2+}$ oxidation state for CMRO and CYMRO (see Section III.A). This is also suggested by previous first-principles electronic structure calculations for CMRO that shows half-filled Mn $3d$ levels with no Mn $3d$ states near the Fermi level \cite{Cavichini2018}. Our XAS derivative and second derivative curves indeed support the same Mn oxidation state for CMRO and CYMRO [see Figures \ref{XANES-Mn}(b) and \ref{XANES-Mn}(c)]. In this context, the substantial differences between the Mn \textit{K} edge XANES spectra of CMRO and CYMRO [see Figure \ref{XANES-Mn}(a)] are ascribed to a possible hybridization between Mn $4p$ and Y $4p$ and/or $5s$ empty states. Such hypothesis is reasonable considering the highly extended character of such unoccupied states.

The Re \textit{L$_3$} edge XANES spectra of CMRO and CYMRO are given in Figure \ref{XANES-Re}(a) together with the ReO$_2$ and ReO$_3$ standards. Both CMRO and CYMRO show stronger white lines with respect to ReO$_2$ and ReO$_3$. For CMRO, the absorption coefficient shows a well defined maximum at 10.5375 keV, whereas for CYMRO a broader structure is observed. The XANES first- and second-derivative curves are given in Figures. \ref{XANES-Re}(b) and \ref{XANES-Re}(c), respectively. For CMRO, the second-derivative curve show a minimum at 10.5375 keV with a shoulder at 10.534 keV, whereas for CYMRO two well defined minima are observed at 10.534 and 10.538 keV. Our results indicate that the XANES white lines of CMRO and CYMRO are composed of two discernible structures separated by $\sim 3.5 - 4.0$ eV. Overall, a spectral weight transfer of the white line of CMRO to lower energies for CYMRO is noticed, being consistent with a reduction of the average Re oxidation state associated with electron-doping. This conclusion is consistent with the augmented Re-O distances for CYMRO with respect to CMRO (see Section III.A).

\section{Discussion}

It is worthwhile to compare the magnetic structures of CMRO \cite{Cavichini2018} and CYMRO. Both structures present the same crystal and magnetic space group $P2_1/n$ and are highly non-collinear, with a FM component along the crystallographic $b$ direction and an AFM component in the $ac$-plane. The magnitude of the Mn/Re magnetic moments are also nearly the same for both compounds [4.35(7)/0.22(4) $\mu_B$ and 4.18(7) /0.20(4) $\mu_B$ for Mn/Re in CMRO and CYMRO, respectively]. Two important differences between these compounds are the distinct $T_c$'s (121 K for CMRO and 150 K for CYMRO) and the relative weight of the FM and AFM components. In fact, at $T=3$ K, the FM component of the Mn moment is 0.8(3) and 3.38(3) $\mu_B$ for CMRO and CYMRO, respectively. Also, the canting angle between Mn neighboring spins in consecutive planes is 109(2)$^{\circ}$ and 59.5(5)$^{\circ}$ for CMRO and CYMRO, respectively. Considering that the structural changes induced by the Y-substitution are rather small,  such significant modifications in the magnetic parameters induced by Y$^{3+}$ substitution on the Ca$^{2+}$ site are not due to steric effects, being actually related to the electron doping that increases the $5d$ electronic population at the Re site (see below).

The Mn-Mn canting angle is temperature-dependent, favoring the FM over the AFM component on warming for both CYMRO (see Figure \ref{Tdep}) and CMRO (Ref. \cite{Cavichini2018}). This indicates that the FM sublattice of the canted magnetic structure is stiffer than the AFM one. This is reasonable considering that the FM component points along the monoclinic principal axis {\it b} (easy-axis) and the AFM sublattice lies in the {\it ac}-plane. It is therefore expected that the excitations of the FM sublattice should be gapped and the in-plane excitations of the AFM sublattice may be gapless, thereby favoring a higher thermal population of the latter at moderate temperatures.

The largely non-collinear magnetic structures of CMRO and CYMRO are intriguing. Re$^{6+}$ and Re$^{5+}$ ions are $5d^1$ and $5d^2$ electronic systems, respectively. The $5d$ levels are split in $t_{2g}$ and $e_g$ levels by octahedral crystal fields. The $t_{2g}$ multiplet acts as an effective $l=1$ moment, which, in the presence of strong spin-orbit coupling, separates in four-fold $J_{eff}=3/2$ and two-fold $J_{eff}=1/2$ levels, where the former is lower in energy. Thus, the Re $5d$ electrons in both CMRO and CYMRO are expected to occupy $J_{eff}=3/2$ states. An important characteristic of such states is that the magnetic moment operator projected in this subspace, $P_{3/2} \mathbf{M} P_{3/2} = 0$ (Refs. \cite{Xiang2007,Lee2007,Chen2010}). Nonetheless, a small magnetic moment parallel to $<\mathbf j>$ is still expected due to the hybridization with $p$ orbitals at the neighboring oxygen sites \cite{Chen2010}. Indeed, the minute Re moments observed in CMRO and CYMRO are consistent with Re $5d$ electrons occupying $J_{eff}=3/2$ states.
In materials containing only $5d$ magnetic atoms in a double perovskite structure, rich phase diagrams including non-collinear magnetic arrangements may occur, arising from anisotropic bond-directional symmetric exchange terms \cite{Chen2010,Dodds2011,Ishizuka2014,Cook2015}. In addition, since $<\mathbf{M}> // <\mathbf{j}>$ for the Re $J_{eff}=3/2$ states, the Re magnetic moment direction is associated with the specific occupied orbital within the $J_{eff}=3/2$ multiplet, leading to electric quadrupolar (orbital) ordering and a possible pathway to magnetoelastic coupling \cite{Chen2010,Dodds2011,Ishizuka2014}.

Another intriguing aspect of the magnetic structures of CMRO and CYMRO is the largely non-collinear arrangements of the Mn$^{2+}$ spins. Indeed, Mn$^{2+}$ ions show half-filled $3d$ shells with $L=0$ and normally give rise to conventional magnetism dictated by a spin Hamiltonian with a dominating symmetric isotropic Heisenberg exchange, from which soft magnetic phases with collinear spin structures are expected. The relatively large $T_c=120-150$ K of CMRO and CYMRO indicates that the Mn-Re nearest-neighbor exchange interaction dominates over second-neighbor Mn-Mn and Re-Re interactions. This $3d-5d$ exchange interaction is likely the reason behind the unusual Mn spin ordering. The observed canting angle between Mn and Re moments in CYMRO is 110(5)$^{\circ}$ for ions located at the same $z$ and 130(20)$^{\circ}$ for ions in consecutive planes. Thus, a possible dominance of directional-dependent anisotropic symmetric exchange or an unusually strong antisymmetric term for the Mn-Re exchange \cite{Popov2003} are possibilities to explain our results.

The temperature dependencies of the lattice parameters of CMRO and CYMRO displayed in Figure \ref{Tdep_lattice} are also interesting. Whereas the undoped sample shows substantial lattice parameter anomalies at the magnetic ordering temperature, such anomalies are nearly vanished under Y-doping. It is plausible to consider that the lattice anomalies below $T_c$ reflect an electric quadrupolar ordering of the Re $5d$ $J_{eff}=3/2$ orbitals. This is reinforced by the increased distortion of the ReO$_6$ octahedra at low temperatures for CMRO \cite{Cavichini2018}. The coincidence of  orbital and magnetic ordering temperatures is a manifestation of the strong Re spin-orbit coupling. The suppresion of the magnetoelastic coupling under electron doping suggests orbital fluctuations induced by the non-integer occupation of Re $5d$ levels. 

\section{Conclusions}

In summary, we investigate the nuclear, magnetic, and local electronic structures of CYMRO and compare with the parent compound CMRO. It is found that the Y-substitution promotes electron doping of the Re-O network whereas keeping the Mn$^{2+}$ oxidation state stable. Such electron doping increases the magnetic ordering temperature, enhances the ferromagnetic component of the largely non-collinear magnetic structure, and destabilizes the Re $5d$ orbital ordering below $T_c$. These observations are discussed in a scenario where the Re $5d$ electrons occupy $J_{eff}=3/2$ levels for which the substantial orbital and spin moments nearly cancel each other, leading to weak Re magnetic moments that are nonetheless able to promote an unusual behavior for the much larger Mn$^{2+}$ magnetic moments via exchange coupling. 

\section{acknowledgments}
The authors thank Rodrigo Pereira for helpful discussions. LLB and LNLS (XAFS1-14567) are acknowledged for concession of beamtime. This work was supported by FAPESP (grant number 2019/10401-9 and 2018/20142-8), CAPES and CNPq (grant number 308607/2018-0 and	409504/2018-1), Brazil.

\maketitle
\section*{References} 
\bibliography{library.bib} 
\maketitle


\appendix{\begin{center}
\large\textbf{Enhanced magnetism and suppressed magnetoelastic coupling achieved by electron doping the spin-orbit Mott insulator Ca$_2$MnReO$_6$}
\end{center}}

\setcounter{section}{0} 

\section{Scanning Electron Microscopy}

\begin{figure}[h]
    \centering
    \includegraphics[width=0.5\textwidth]{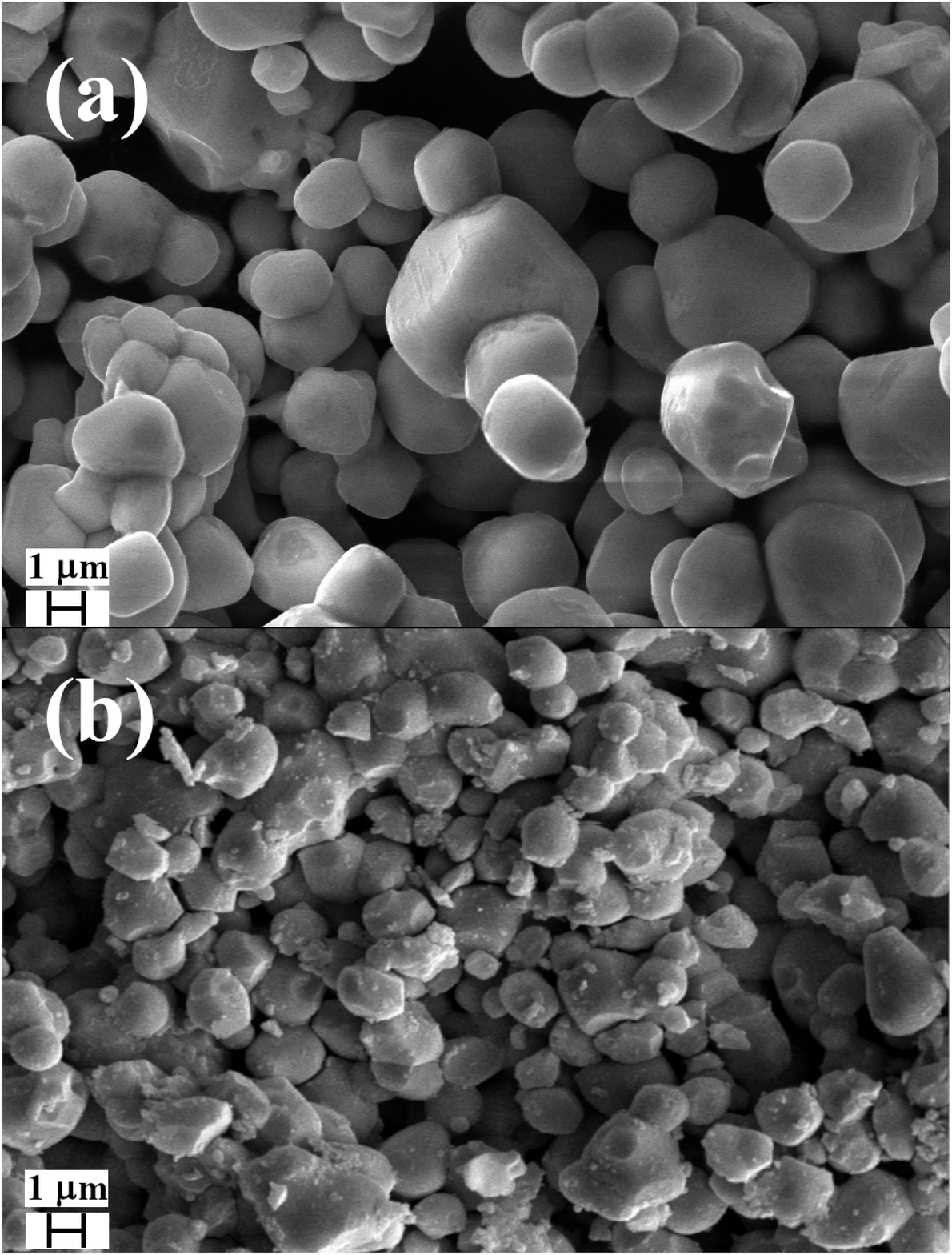}
    \caption{\label{SEM} Secondary electron images of Ca$_2$MnReO$_6$ (a) and Ca$_{1.7}$Y$_{0.3}$MnReO$_6$ (b).}
\end{figure}

In order to observe the sample morphology of our Ca$_2$MnReO$_6$ and Ca$_{1.7}$Y$_{0.3}$MnReO$_6$ samples, Scanning Electron Microscopy (SEM) images were obtained with a ZEISS EVO-40 microscope. SEM image (Figure \ref{SEM}) evaluations using the Fiji software \cite{Fiji} lead to average particle diameters of 2.2(3) $\mu$m for Ca$_2$MnReO$_6$ and 1.5(2) $\mu$m for Ca$_{1.7}$Y$_{0.3}$MnReO$_6$.

\section{Thermal neutron powder diffraction}

\begin{figure}[ht]
\centering
\includegraphics[width=1\textwidth]{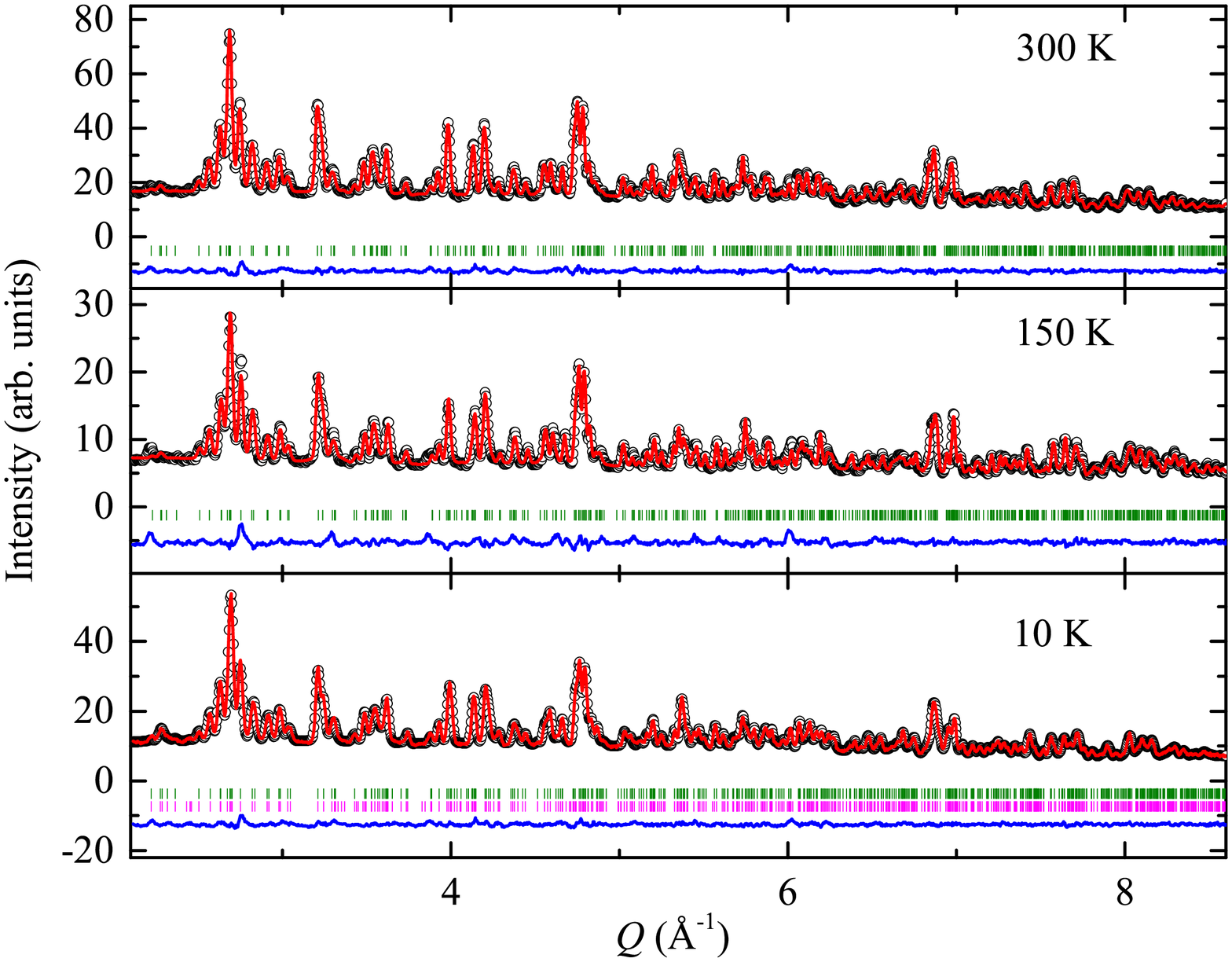}
\caption{\label{t-npd} Observed (open symbols) and calculated (red solid lines) thermal neutron powder diffraction profiles of Ca$_{1.7}$Y$_{0.3}$MnReO$_6$ at $T=300$, 150 and 10 K. The difference curves (blue solid line) are displayed at the bottom of each figure. The reflection positions are indicated as short vertical bars for the nuclear structure at all temperatures (green) and for the magnetic structure at $T=10$ K (magenta). }
\end{figure}

In order to obtain accurate structural parameters of Ca$_{1.7}$Y$_{0.3}$MnReO$_6$, high-resolution thermal neutron powder diffraction (t-NPD) measurements were taken at $T=300$, 150 and 10 K at the $3T2$ instrument of LLB using a vertically focusing Ge(335) monochromator with $\lambda = 1.2292$ \AA\ and a bank of 50 $^{3}$He detectors. The observed and calculated diffraction data are shown in Figure \ref{t-npd}. The refined structural parameters and bond distances are given in Tables \ref{struct} and \ref{bonds} of the main text.

\section{X-ray powder diffraction}

\begin{figure}[ht]
\centering
\includegraphics[width=1\textwidth]{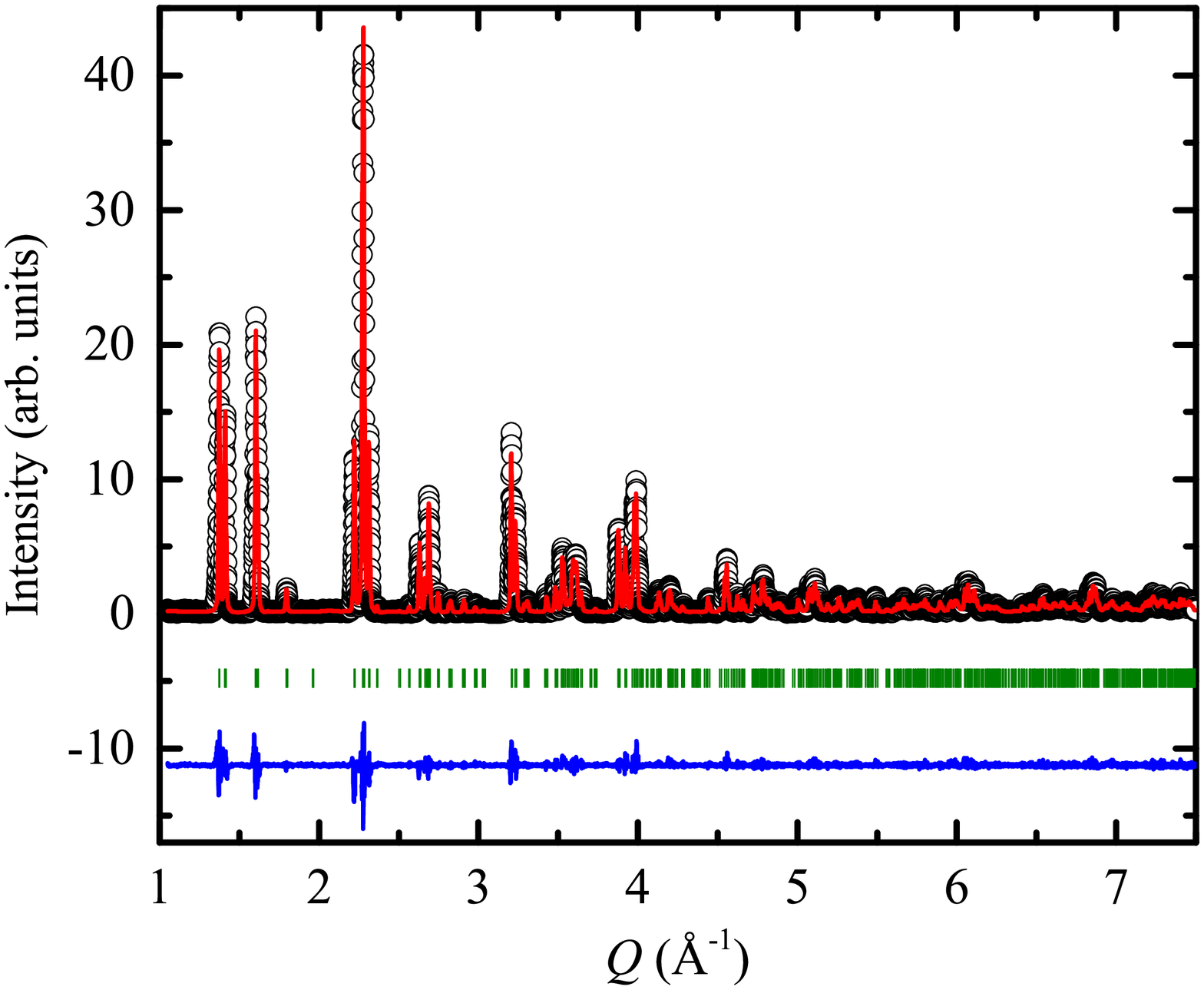}
\caption{\label{XRD} Observed (open symbols) and calculated (red solid lines) X-ray powder diffraction profile of Ca$_{1.7}$Y$_{0.3}$MnReO$_6$ at room temperature. The difference curves (blue solid line) are displayed at the bottom of each figure. The reflection positions are indicated in short vertical bars.}
\end{figure}

X-ray powder diffraction measurements of Ca$_{1.7}$Y$_{0.3}$MnReO$_6$ were performed on a Rigaku Ultima IV diffractometer with Bragg-Bretano geometry equipped with a lithium flouride (LiF) using CuK$\alpha$ radiation (see Figure \ref{XRD}), supporting full chemical ordering at the Mn/Re transition-metal sites of the double perovskite structure. This conclusion is in line with an independent analysis using neutron powder diffraction data (see main text).  

\section{{\it ac}-magnetic susceptibility}

{\it ac}-Susceptibility measurements were taken with frequency $f=448$ Hz and magnetic field strength amplitude $H_{ac}=7.5$ Oe in a home-made {\it ac} susceptometer such as described in Ref. \cite{BOrlando2019} using a Stanford Instruments SR850 lock-in amplifier and a Hewlett-Packard HP 3440A voltmeter.

\begin{figure}[t]
\centering
\includegraphics[width=1\textwidth]{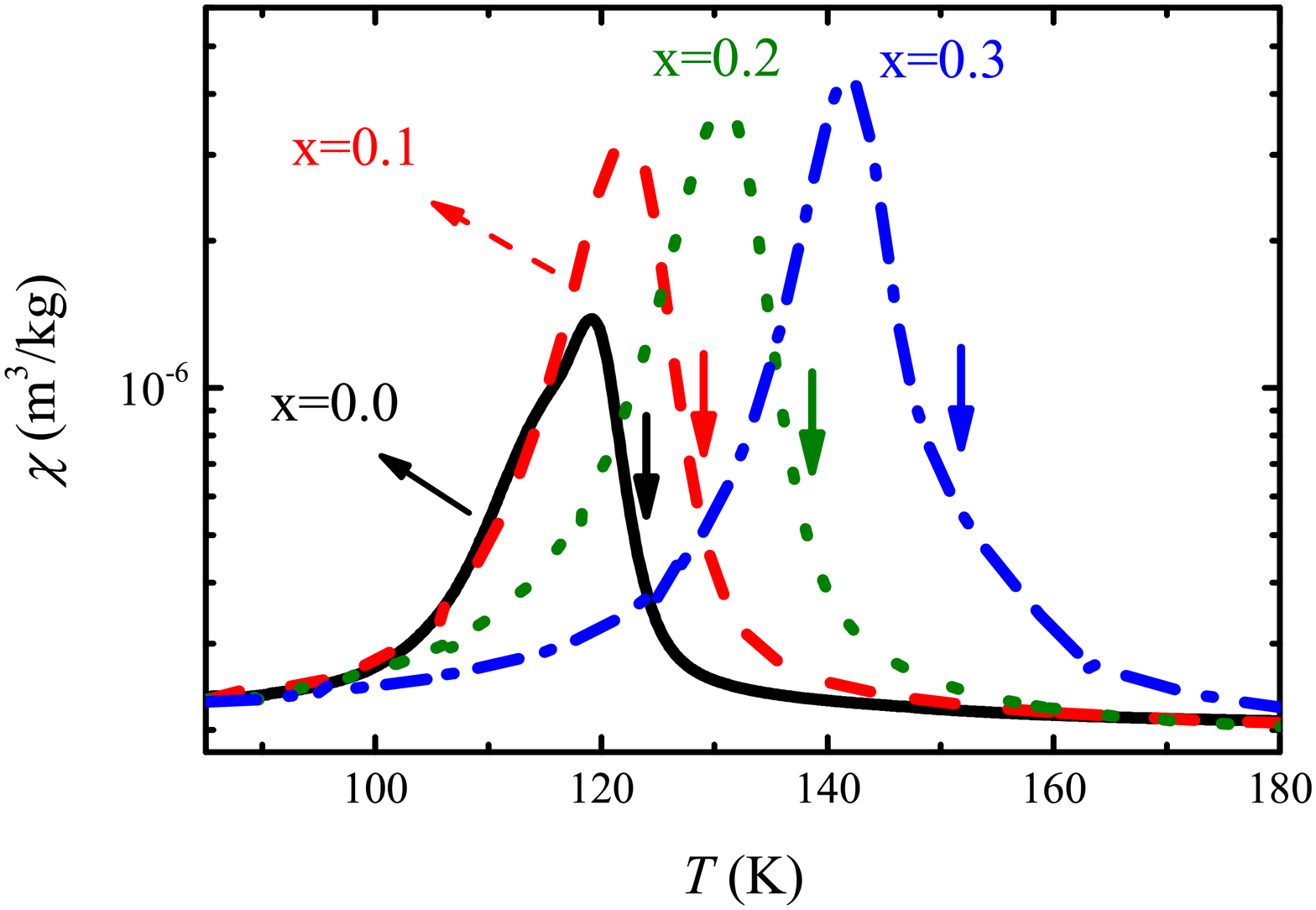}
\caption{\label{chiacXT} {\it ac}-Magnetic susceptibility curves for Ca$_{2-x}$Y$_x$MnReO$_6$ with $x=$0.0, 0.1, 0.2, and 0.3. The estimated magnetic ordering transition temperatures are indicated by vertical arrows.}
\end{figure}

Figure \ref{chiacXT} shows the {\it ac}-magnetic susceptibility $\chi(T)$ of Ca$_{2-x}$Y$_x$MnReO$_6$ with $x=0.0$, 0.1, 0.2, and 0.3 (sample $1$), taken on cooling. The corresponding magnetic ordering temperatures $T_c$ are identified with arrows. A continuous increment of $T_c$ with $x$ is observed, ranging from $T_c=121$ K for $x=0.0$ (CMRO) to $T_c=150$ K for $x=0.3$ (CYMRO). For all investigated samples, $\chi(T)$ present a maximum at a temperature slightly below $T_c$ and shows a strong reduction below such temperature.  This behavior is likely related to the freezing of the magnetic domains in the magnetically ordered phase.

\section{{\it dc}-electrical resistivity}

\begin{figure}[ht]
\centering
\includegraphics[width=1\textwidth]{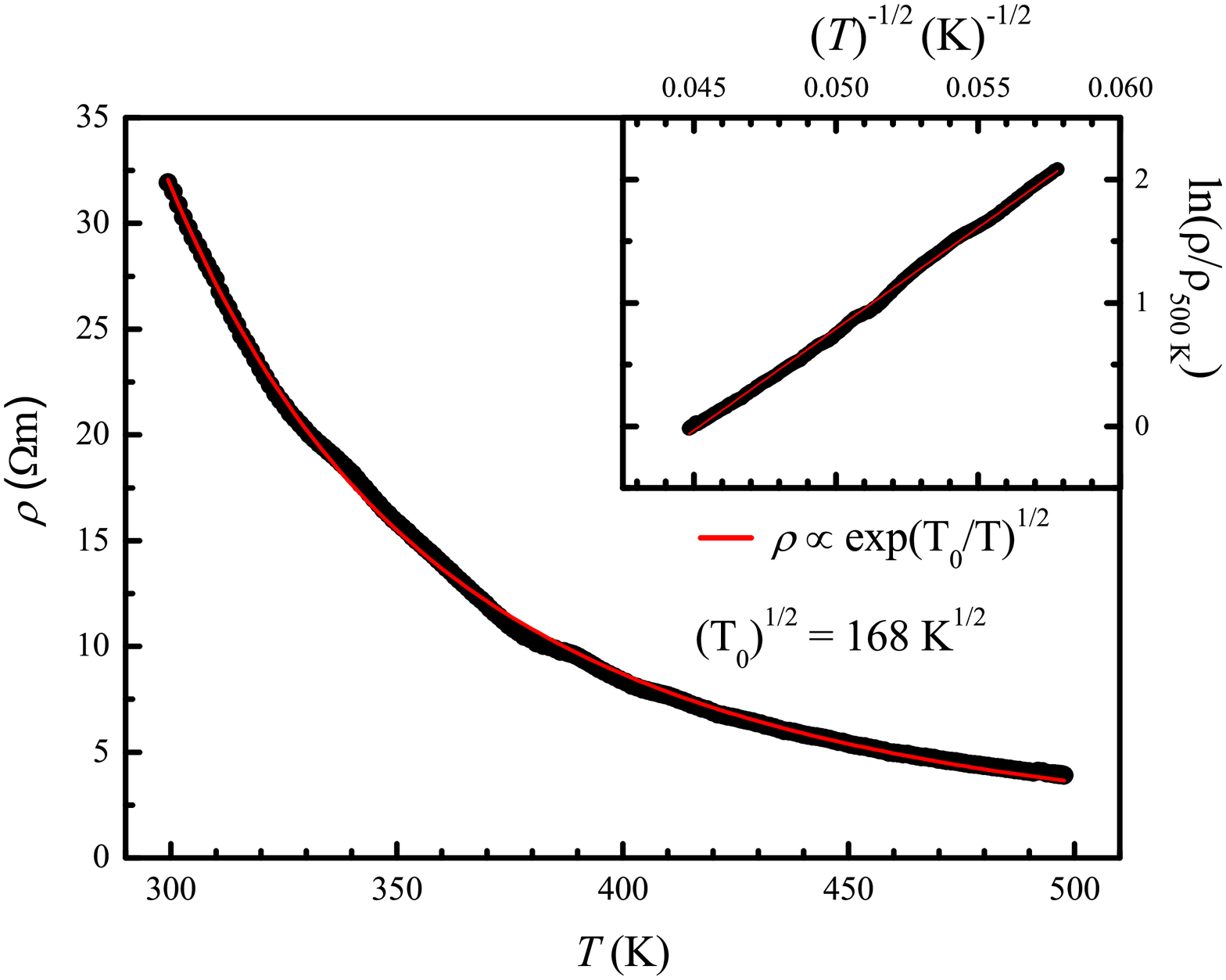}
\caption{\label{Res} Resistivity as a function of temperature for CYMRO (sample $2$) with an exponential fitting (red line) for a $\rho \propto exp{(T_{0}/T)^{1/2}}$ function (ES-VRH model, see text). The inset panel shows the plot of ln$(\rho/\rho_{(500 K)})$ versus $T^{-1/2}$, where $\rho_{(500 K)}$ is the resistivity measured at 500 K.}
\end{figure}

The electrical resistivity of Ca$_{1.7}$Y$_{0.3}$MnReO$_6$ was measured on a thin piece of a pelletized sample by the standard four-probe method using a Keithley 2182A nanovoltmeter and a Keithley 6221 current source, with an electrical current $I = 20$ mA. The sample was mounted at the cold finger of a closed-cycle He cryostat, and thermal contact was ensured by a standard thermal paste, which also prevented current leaks. Temperature-dependent electrical resistivity of Ca$_{1.7}$Y$_{0.3}$MnReO$_6$ is shown in Figure \ref{Res}. A non-metallic behavior is observed in the investigated temperature range ($300 < T < 500$ K), where $\rho \propto exp{(T_{0}/T)^{1/2}}$. The fitting parameter $(T_{0})^{1/2} = 168$ K$^{1/2}$ is smaller than the mean value previously reported for the pure compound $(T_{0})^{1/2} = 277$ K$^{1/2}$ (Ref. \cite{Fisher2008}).

The similar non-metallic character of CYMRO and CMRO (see Figure \ref{Res} and Ref. \cite{Fisher2008}) indicates that the electron-doping associated with the Y substitution is not sufficient to turn this material into a metal, and the electron conduction remains thermally activated. It is possible that the density of states at the Fermi level remains critically low for CYMRO, preventing a metallic behavior. Many correlated insulating materials are characterized to obey Mott's law $\rho \propto exp{(T'_{0}/T)^{1/4}}$ (Ref. \cite{MOTT19681}), where the electron transport h
ns via variable-range hopping (VRH) mechanism. Another key ingredient that leads to this temperature dependence is the behavior of the density of states (DOS) near the Fermi level, which in this case is assumed to be constant. Based on Motts's VRH mechanism, Efros and Shklovski (Ref.  \cite{Efros_1975}) proposed a related model (ES-VRH), assuming however that the DOS is parabolic near the Fermi level, which gives rise to $\rho \propto exp{(T_{0}/T)^{1/2}}$. In both cases, $T'_{0}$ and $T_{0}$ are characteristic temperatures related to the dielectric constant and the spatial length of the localized electronic wave function. For our fits for CYMRO, the exponent converged to 0.5, being subsequently kept fixed, indicating that the electrical transport mechanism of CYMRO is well described by the ES-VRH \cite{Efros_1975} approach.

\end{document}